\documentclass[aps,prl,showpacs,reprint,superscriptaddress]{revtex4-1}
\usepackage{graphicx,color,epsfig}
\usepackage{amsmath,amssymb,bm}

\begin{document}
\title{Leaders and obstacles raise cultural boundaries}
\author{M. G. Cosenza}
\affiliation{School of Physical Sciences \& Nanotechnology, Universidad Yachay Tech, Urcuqu\'i, Ecuador}
\author{O. Alvarez-Llamoza}
\affiliation{Grupo de Simulaci\'on, Modelado, An\'alisis y Accesabilidad, Universidad Cat\'olica de Cuenca, Cuenca, Ecuador}
\author{C. Echeverria}
\affiliation{CeSiMo, Facultad de Ingenier\'ia, 
Universidad de Los Andes, M\'erida, Venezuela}
\author{K. Tucci}
\affiliation{CeSiMo, Facultad de Ingenier\'ia, 
Universidad de Los Andes, M\'erida, Venezuela}

\date{October 2020}

\begin{abstract}
We employ an agent-based model for cultural dynamics to investigate the 
effects of spatial heterogeneities on the collective behavior of a social system.   
We introduce
heterogeneity as a random distribution of defects or imperfections in a two-dimensional lattice. Two types of defects are considered separately: obstacles that represent geographic features, 
and opinion leaders, described as agents exerting unidirectional influence on other agents. In both cases,
we characterize two collective phases on the
space of parameters of the system, given by the density of defects and a quantity
expressing the number of available states: one ordered phase, consisting of a large homogeneous group; and a disordered phase, where
many small cultural groups coexist. In the case of leaders, the homogeneous state corresponds to their state.
We find that 
a high enough density of obstacles contributes to cultural diversity in the system. On the other hand, we find a nontrivial effect when opinion leaders are distributed in the system:  if their density is greater than some threshold value, leaders are no longer efficient in
imposing their state to the population, but they actually promote multiculturality. In this situation, we uncover that leaders, as well as obstacles, serve as locations for the formation of boundaries and segregation 
between different cultural groups. 
Moreover, a lower density of leaders than obstacles is needed to induce  multiculturality in the system. 
\end{abstract}

\maketitle


\section{Introduction}
The study of dynamical processes on nonuniform or heterogeneous media
is a topic of wide interest.  
The nonuniformity may arise from the
intrinsic heterogeneous nature of the substratum, 
such as porous or fractal media, or it may be due to
random defects or vacancies in the medium at some
length scales \cite{Avnir}, or it may consist of added impurities in a material, as in semiconductor doping \cite{Paul}. 
Heterogeneities can have significant effects in the behavior of a system;
for example, imperfections in the crystal lattice of a solid can produce changes in properties 
that open up new applications \cite{Krystyn}, or they  can affect the formation of spatial patterns, 
by inducing phase transitions in excitable media \cite{Clar}, and defects can serve as nucleation sites for growth processes \cite{Puri}. 

In the context of Social Sciences, the relation between
spatial heterogeneities, in the form of geographic features, and the development and dissemination of cultures,
has been an issue of long standing interest \cite{Sauer,Gregory}. Recently,
technological resources have allowed the study of 
geographically embedded networks \cite{Ozga}, or
the relationship between geographic constraints and social network communities
\cite{Onnela}. Spatial inhomogeneities in social dynamics may also occur by the presence of
elements such as influential agents, distributed mass-media messages, or outdoor advertising \cite{Katz,Valente,Roch}. 

In this article we employ an agent-based model to investigate the 
effects of spatial heterogeneities on the 
collective behavior of a social system.  We introduce
heterogeneity as defects or imperfections in a two-dimensional lattice. 
In Section~II, we present a general model for cultural dynamics on a lattice with distributed defects.
In a first application, defects correspond to vacancies that simulate landscape obstacles. In a second model, we associate defects to influential agents distributed on the lattice. 
Influentials are considered as a minority of agents who can influence a large number of
people, but are
less susceptible to influences from other agents
 \cite{Aral,Gadwell}.
In opinion formation models,
these agents are named opinion leaders or spreaders \cite{Tucci,Weimann,Lloyd,Borge}; they are also called inflexibles, zealots, or committed agents 
\cite{Mobilia,Galam,Verma,Xie}. 
We employ the spatial density of defects --either obstacles or opinion leaders-- as a parameter of the system. The interaction dynamics is based on Axelrod's rules for the dissemination of culture among social agents \cite{AL1}, a paradigmatic model of much interest in Sociophysics 
\cite{Schweitzer,CMV1,KETS2,KETS1,Kuperman,VVC1,Santis,Gracia,Zhang,GCT1,Peres}.

Section~III contains our main results. We find that 
a high enough density of obstacles induces cultural diversity in the system. On the other hand, we encounter a nontrivial effect when opinion leaders are distributed in the system:  if their density is greater than some threshold value, leaders are not efficient in
imposing their message to the population, but they actually promote multiculturality. We find that leaders, as well as obstacles, serve as locations for the formation of boundaries and segregation 
between different cultural groups. 
Moreover, it takes a lower density of leaders than obstacles to induce  multiculturality in the system. Section~IV presents the Conclusions of this work.

\section{Cultural dynamics on lattices with defects}
An inhomogeneous lattice can be generated as follows. Start from a two-dimensional array of sites of size $N=L \times L$ with periodic boundary conditions, and set as defects a given fraction $\rho$ of sites at random. 
The distribution of defects on the lattice
can be characterized in terms of the minimum Euclidean distance $d$ between defects, as shown in Fig. 1. With this setting,
the density of defects scales with $d$ as 
$\rho= 0.625d^{-2}$ \cite{Feder,Impure}. 

\begin{figure}[h]
\includegraphics[scale=0.7,angle=0]{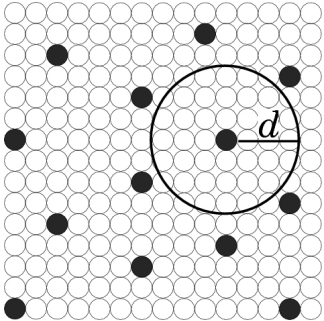}
\caption{Two-dimensional spatial support of the system, showing active sites ($\circ$) and defects ($\bullet$). Defects are randomly placed in such a way that their density distribution is maximum for a given minimum distance $d$ between them. 
\label{fig:1}}
\end{figure}

There are $\rho N$ randomly distributed defects in the lattice.
The remaining $N(1-\rho)$ sites consist of \textit{active} agents, susceptible of changing their states.
Following Axelrod's model \cite{AL1},
the state of an active agent $i$ $(i=1,2,\ldots,N(1-\rho))$ is given by the $F$-component vector $c_i=(c_i^1, \ldots,c_i^f, \ldots, c_i^F)$,
where each component $c_i^f$ represents a cultural feature that can take any of $q$ different traits or options in the set $\{0, 1, . . . , q-1\}$. There are $q^F$ possible equivalent states. 

In a first model, the $\rho N$ randomly distributed defects correspond to obstacles or vacancies in the lattice.
We define obstacles as empty sites that possess no dynamics; they may represent dispersed geographic features.
In the second model,
 we  consider that
the $\rho N$ randomly distributed defects are occupied by
opinion leaders or influential agents. We assume that opinion leaders share the same fixed state, denoted by
$(y^1, \ldots,y^f, \ldots, y^F)$,
where each component
$y^f \in \{0, 1, . . . , q-1\}$ remains unchanged during the evolution of the system. Additionally,
we assume that the interaction of opinion leaders with active agents is unidirectional; i.e.,
opinion leaders can affect the state of other agents, but their state does not change. This simplifying assumption reflects the basic role attributed to
opinion leaders and influentials in a society  \cite{Katz,Valente,Roch,Aral,Gadwell,Tucci,Weimann,Lloyd,Borge,Mobilia,Galam,Verma,Xie}. This also comprises the notion of cultural
status proposed by Axelrod \cite{AL1}. 
With these conditions, opinion leaders can be seen as
a spatially nonuniform mass media field acting on the system, or as distributed sources-transmitters of mass media messages.

As initial condition in both models, each active agent is randomly assigned one of the $q^F$ possible states with a uniform probability. In this article, we fix $F=10$ as the number of cultural features.
We define 
the dynamics of
the system for both models, in the presence of leaders (obstacles), by iterating the following steps:
 
\begin{enumerate}
\item Select at random an active agent $i$.
\item Select at random an agent $j$ among the eight nearest neighbors of $i$
in a Moore's neighborhood (such as $j$ is not an obstacle).
\item Calculate the \textit{overlap} between the state $c_i$ of active agent $i$ and the state $c_j$ of its selected neighbor $j$, given by the number of shared cultural features between their respective vector states, as follows
      \begin{equation}
  l(i,j)=    \left\lbrace 
 \begin{array}{ll}  
\sum_{f=1}^F \delta_{c_i^f \, c_j^f}, & \mbox{if $j$ is an active agent},\\
\sum_{f=1}^F \delta_{c_i^f \, y^f},   & \mbox{if $j$ is a leader}. 
  \end{array}
     \right.  
\end{equation}
The delta Kronecker function employed is  
    $\delta_{x,y} = 1$, if $x=y$; $\delta_{x,y} = 0$, if $x\neq y$.
\item If $0 < l(i,j) < F$, with probability $l(i,j)/F$ choose $h$ randomly such that $c_i^h \neq c_j^h$ and set 
$c_i^h=c_j^h$ if $j$ is an active agent,  
or $c_i^h=y^h$ if $j$ is a leader. 
If $l(i,j)=0$ or $l(i,j)=1$, the state $c_i$ does not change.     
\end{enumerate}

In the absence of defects $(\rho = 0)$, a system subject to Axelrod's dynamics reaches a stationary configuration
in any finite network, where the agents constitute domains of different sizes. A domain is a set of connected agents that share
the same state. 
A homogeneous or ordered phase in the system is characterized by 
the presence of one domain.
The coexistence of several domains
corresponds to an inhomogeneous or disordered phase in the system. On different networks, the system undergoes a nonequilibrium transition
between
an ordered phase for values $q < q_o$, and a disordered phase for $q > q_o$, where $q_o$ is a critical value \cite{CMV1,KETS2}.
The critical value for $F = 10$ on a two-dimensional lattice  has been numerically estimated at $q_o \approx 55$ \cite{KETS1}.

\section{Influence of opinion leaders and obstacles in cultural dynamics}

\begin{figure}[h]
\centerline{\includegraphics[scale=0.34,angle=0]{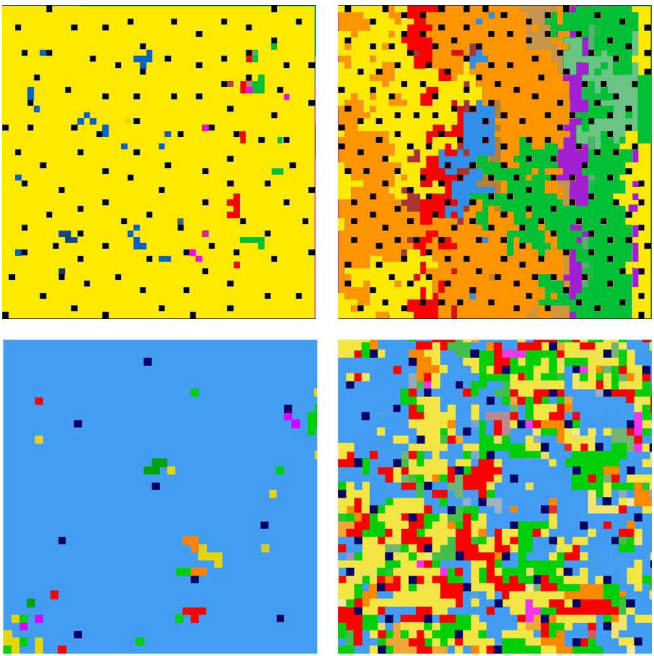}}          
\caption{Stationary spatial patterns for a two-dimensional lattice with different densities of obstacles (top panels) 
and leaders (bottom panels). Fixed parameters: $N=50 \times 50$, $F=10$, $q=44<q_o$. Different domains are represented by different colors.
Top: obstacles are indicated as black sites with densities $\rho=0.04$ (left) and $\rho=0.14$ (right).
Bottom: leaders are identified as black sites for visualization, but their fixed state corresponds to the light blue predominant color on the left panel, $\rho=0.005$ (left); $\rho=0.04$ (right).}
\label{fig:2}
\end{figure}

When defects in the form of leaders or obstacles are present in the lattice, the order-disorder transition persists, but the
critical value $q_c$ for which the transition occurs decreases as the density $\rho$ is increased. Figure~\ref{fig:2}
shows the spatial configurations of the asymptotic stationary states of the
system with different densities of leaders or obstacles, 
and fixed $q=44<q_o$.

The patterns  corresponding to the model with obstacles are exhibited in the top panels of Fig.~\ref{fig:2}. We observe that the system reaches a homogeneous state when the density of obstacles is small. The homogeneous state can be any of the possible $q^F$ states, depending on initial conditions. Increasing the density of obstacles above some threshold value leads to the formation of multiple domains or multiculturality.
The bottom panels of Fig.~\ref{fig:2} show the behavior of the model with opinion leaders.
When the density of opinion leaders is small, the system is driven
towards a homogeneous state equal to the state of the leaders. However, as the density of leaders is increased, the system no longer
converges to the state of the leaders, but reaches a disordered state. Domains having a cultural state
equal to that of the opinion leaders survive, but they become smaller in size as $\rho$ increases. Thus,
we have the counterintuitive effect that, above some threshold
value of their density, opinion leaders actually induce cultural diversity in the system.

\begin{figure}[h]
 \includegraphics[width=0.33\textwidth,angle=90]{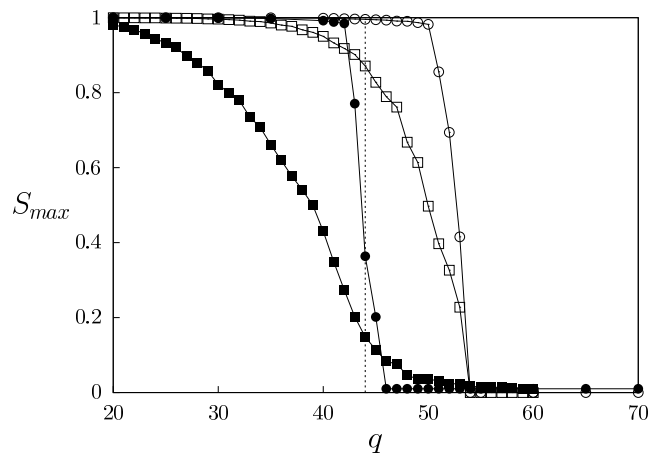}
 \caption{Order parameter $S_{max}$ as a function of $q$, for
 different values of $\rho$, with  $F=10$ and size $N=100 \times 100$. The values $S_{max}$ are averaged over $50$ realizations of initial conditions for each value of $q$.  Curves for the model with obstacles correspond to  $\rho=0.04$ (empty squares) and $\rho=0.14$ (solid squares), while curves for the model with leaders are plotted for $\rho=0.005$ (empty circles) and $\rho=0.04$ (solid circles).
 The dotted vertical line signals the value $q=44$ used in Fig.~\ref{fig:2}.}
 \label{fig:3}
 \end{figure}

A useful order parameter to characterize the collective behavior is the average normalized size of the largest domain in the system \cite{CMV1}, denoted by $S_{max}$.  Here, we calculate the quantity $S_{max}$
over the set of active agents in the system, of size $N(1-\rho)$. 
Figure~\ref{fig:3} shows $S_{max}$ as a function of the number of options $q$ for both models, with leaders or obstacles, for the values of density employed in Fig~\ref{fig:2}. 
For each value
of the density of leaders $\rho$, we observe a transition at a critical value $q_c(\rho)$, from
the homogeneous cultural state imposed by the leaders,
where $S_{max} = 1$, to a disordered or multicultural state for
which $S_{max} \to 0$. 
The critical value $q_c(\rho)$ decreases
with increasing $\rho$. 
Similarly, in the obstacle model, 
an order-disorder transition takes place at a critical value of $q$ that decreases as the density of obstacles is increased.

\begin{figure}[h]
\includegraphics[width=0.32\textwidth,angle=90]{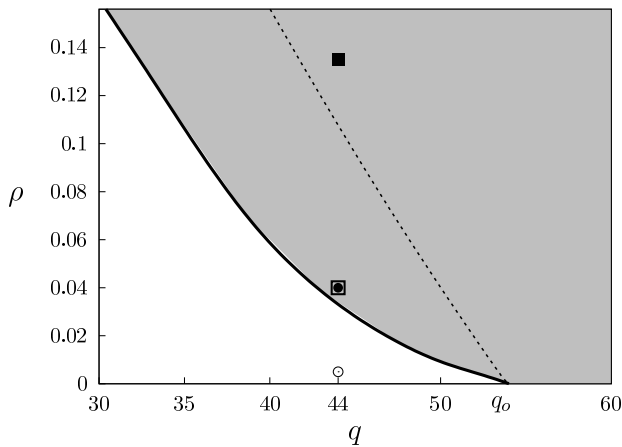}
\caption{Phase diagram for the system with leaders and with obstacles.
 Fixed parameters are  $F=10$, $N=100 \times 100$.
 Leaders: the continuous black line is the critical boundary separating the ordered, homogeneous phase (white color) from the disordered phase (gray color). Obstacles: the dotted line separates the homogeneous phase (below the line)  from the disordered phase (gray color, above the line).
The marked points correspond to $q=44$ and the density values used in
Fig.~\ref{fig:3}: obstacles with $\rho=0.04$ (empty square);
obstacles with $\rho=0.14$ (solid square);
 leaders with $\rho=0.005$ (empty circle); leaders with $\rho=0.04$ (solid circle).}
\label{fig:5}
\end{figure}

Figure~\ref{fig:5} shows the phase diagram of the system for both models,
with leaders and with obstacles,
on the space of parameters $(\rho,q)$. 
There is a critical curve 
$q_c(\rho)$ in each case, signaled by a continuous line for the model with leaders and by 
a dotted line for the system with obstacles. Each critical curve separates two well defined collective phases: an ordered or homogeneous phase below the curve, characterized by $S_{max} = 1$; and a disordered or multicultural phase above the curve, where $S_{max} \to 0$. For the model with leaders, these phases are indicated in Fig.~\ref{fig:5} by the regions 
in white and gray colors, respectively.
In both models, when $\rho=0$, we recover the critical value $q_o$ corresponding to the transition in Axelrod's model.

Below the critical boundary indicated by the dotted line in Fig.~\ref{fig:5}, the homogeneous phase reached in the model with obstacles can be any of the possible states in the system. 
A high enough density of obstacles leads to multiculturality or a disordered phase, a result that one may expect, since geographic obstacles usually
contribute to the separation of cultures \cite{Gregory}. Counterintuitively, if the density of opinion leaders is greater than some critical value, leaders are not longer efficient in
imposing their message to the population of active agents, and disorder ensues. 
Thus, leaders act as obstacles on the collective behavior of the system when their density exceeds a threshold value.
Moreover, for a given number of options $q$, a lower density of leaders than obstacles is required to produce multiculturality in the system.

To elucidate the mechanism by which leaders and obstacles induce
multiculturality, 
we consider 
the subset of sites that constitute the interface between different cultural domains in the stationary configuration of the system.
We define that a site belong to the interface subset if
at least three of its neighbors share its state, and at least three other share a different state. Then, we calculate the 
fraction of those sites in the interface that are leaders (or obstacles), denoted by $\rho_f$. 
A ratio value $\rho_f/\rho < 1$ indicates that the density of leaders (obstacles) lying on the 
interface is lesser than the density of leaders (obstacles) in the entire lattice. This 
means that, proportionally, more leaders (obstacles) remain in the interior of the domains than on the interface between domains. When $\rho_f/\rho > 1$, the opposite situation takes place: leaders (obstacles) tend to lie on the interface between domains rather than inside the domains.

\begin{figure}[h]
\includegraphics[width=0.68\linewidth,angle=90]{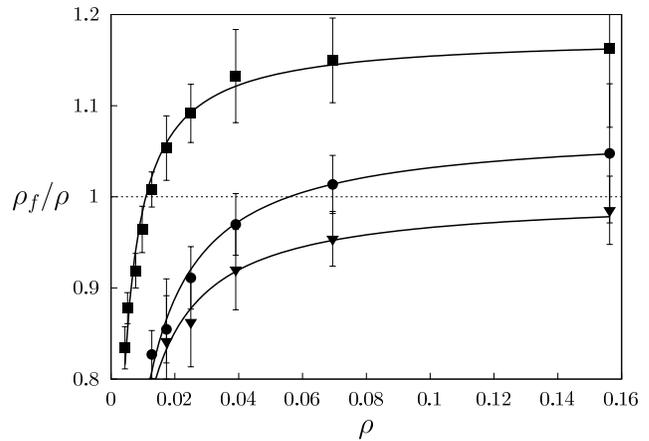}
\caption{Mean ratio $\rho_f/\rho$ as a function of the density of leaders $\rho$, for different values of $q$, calculated over $30$ realizations. Errors bars indicate $\pm 1$ standard deviations. 
The dotted horizontal line signals the value $\rho_f/\rho=1$.
Curves shown for: $q=30$ (triangles); $q=40$ (circles);  $q=50$ (squares).}
\label{fig:rhoFrontera}
\end{figure}

Figure~\ref{fig:rhoFrontera} shows the mean ratio $\rho_f/\rho$ as a function of the density of leaders in the system $\rho$, for different
values of $q$, calculated over several realizations of initial conditions.
We observe that the value of the ratio $\rho_f/\rho$ increases as the density of leaders $\rho$ is incremented. For $q=30$, the ratio
$\rho_f/\rho<1$ on the range of values of $\rho$ displayed.  
On the other hand, 
the phase diagram in Figure~\ref{fig:5} shows that, for that range of values of $\rho$
and $q=30$, the system always reaches the homogeneous state imposed by the leaders. For both values $q=40$ and $q=50$ in 
Fig.~\ref{fig:rhoFrontera}, we see a change in the ratio, from $\rho_f/\rho < 1$ to 
$\rho_f/\rho > 1$, as $\rho$ varies. Correspondingly, for the parameter values $q=40$ and $q=50$, Fig.~\ref{fig:5} reveals a change of the collective phase of the system, from homogeneous to disordered,
as $\rho$ increases across the critical boundary.

\begin{figure}[h]
\includegraphics[width=0.68\linewidth,angle=90]{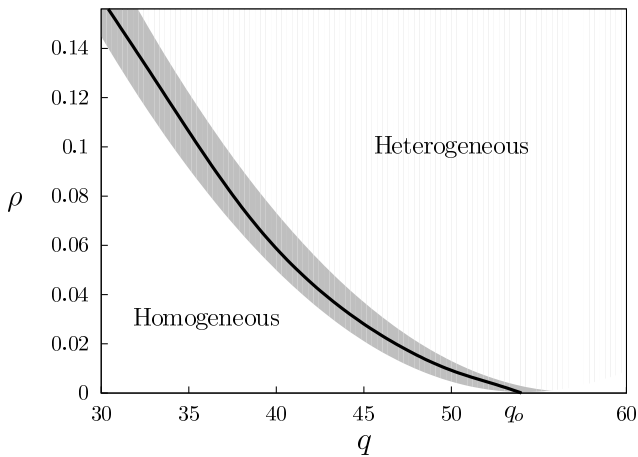}
\caption{Values of $\rho$ for which $\rho_f/\rho=1$ as a function of $q$, within statistical deviations indicated by a gray band. The continuous black curve corresponds to the critical boundary that separates the homogeneous from the heterogeneous phases for the system with leaders in Fig.~\ref{fig:5}.}
\label{fig:rhoFrontera2}
\end{figure}

These results suggest that the ordering properties of the system under the influence of leaders are related to the ratio $\rho_f/\rho$.  To investigate this relation, we plot in Fig.~\ref{fig:rhoFrontera2}
the value of $\rho$ for which $\rho_f/\rho=1$ as a function of $q$. Because of the error bars in the quantity $\rho_f/\rho$ shown in Fig.~\ref{fig:rhoFrontera}, this value of $\rho$ is determined within an error interval, displayed in  Fig.~\ref{fig:rhoFrontera2} as the vertical width in gray color for each value of $q$. The critical boundary $q_c(\rho)$, that separates the homogeneous from the disorder phases
for the system with leaders, is also shown. This boundary coincides, within statistical fluctuations, with the 
condition $\rho_f /\rho=1$ as a function of $q$. 
Thus, we find that the homogeneous phase is associated to the condition $\rho_f /\rho<1$, where leaders more likely lie inside the domains.  On the other hand, the disordered phase
corresponds to the relation $\rho_f /\rho>1$ expressing that leaders are mostly concentrated along the boundaries between cultural domains. In other words, when the density of leaders is greater that some threshold value, 
leaders change their function: they no longer serve as nucleation sites for the growth of their cultural domains, but become 
locations that facilitate the formation of the interface in the system. As a consequence, leaders contribute to the emergence of multiculturality. Since leaders siting on a boundary must have neighbors that share their state, leaders promote the containment and segregation of their followers from other cultural domains.

\begin{figure}[h]
\includegraphics[width=0.68\linewidth,angle=90]{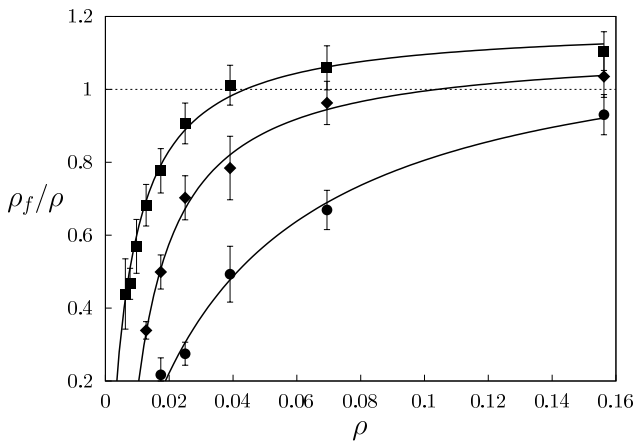}
\caption{Mean ratio $\rho_f/\rho$ as a function of the density of obstacles $\rho$, for different values of $q$, calculated over $30$ realizations. Errors bars indicate $\pm 1$ standard deviations. 
The dotted horizontal line marks the value $\rho_f/\rho=1$.
Curves shown for: $q=40$ (circles); $q=44$ (diamonds); $q=50$ (squares).}
\label{fig:rhoFronteraDefectos}
\end{figure}

Figure~\ref{fig:rhoFronteraDefectos} shows the ratio $\rho_f/\rho$ as a function of $\rho$, for several values of $q$, on a lattice with obstacles. 
Again, the condition $\rho_f /\rho>1$, that characterize the prevalence of obstacles on the interface rather than inside domains, is related to the multicultural phase. 
We have verified that
the critical curve (dotted line) on the phase diagram of Fig.~\ref{fig:5} corresponds to the value of the density of obstacles $\rho$,  for which $\rho_f /\rho=1$, as a function of $q$ within statistical deviations. Then,
obstacles induce multiculturality by facilitating the occurrence of boundaries along their sites that separate different domains. 

Our results reveal that, when the concentration of opinion leaders is sufficiently large,  they actually behave as obstacles: leaders favor the formation of cultural boundaries in a social system
on a two-dimensional space. 
We have verified that employing a square lattice with 
free boundary conditions as spatial support or increasing the system size do not
 significantly change these results. 

\section{Conclusions}

We have employed an agent-based model for cultural dynamics to investigated the 
effects of spatial heterogeneities on the collective behavior of a social system.  Heterogeneities are introduced as a random distribution of defects or impurities 
in a two-dimensional lattice. We have employed the spatial density of defects as a relevant parameter for the collective behavior of the system. We have considered two types of defects: 
(i) obstacles or vacancies that represent geographic features; and (ii) opinion leaders as mediators-originators of mass media messages or advertisement.

A high enough density of obstacles contributes to cultural diversity in the system.
This result should be expected, since geographic features such as mountains, lakes, and deserts, have traditionally restricted cultural exchanges \cite{Sauer,Gregory}.
In the contemporary globalized world, landscape obstacles have become irrelevant.
However, some ecosystems with natural barriers such as tropical jungles still harbor uncontacted cultures \cite{Clark}. Our model with opinion leaders shows a counterintuitive effect: a sufficiently large concentration of opinion leaders, far from favoring the spreading of their message to the entire system, actually promotes multiculturality. This effect is similar to that produced by a global uniform mass media field acting on a regular lattice when the intensity of the field is increased \cite{GCT1}. The model with distributed opinion leaders can be considered as a system subject to a spatially nonuniform field.  Thus, the density of leaders is analogous the intensity of an external global field when considering the collective dynamics of a social system.

We have investigated the mechanism by which obstacles and opinion leaders induce cultural diversity when their density increases.  Above some respective critical value of their
density, obstacles as well as leaders, serve as locations for the formation of the interface or boundaries
between different cultural domains. A lower density of leaders than obstacles is needed to induce  multiculturality in the system. 
In particular, leaders change their role when their density crosses over a critical value, from mostly being efficient transmitters of their message inside a domain, to become sites that rather facilitate the emergence of cultural boundaries.

Our results should be relevant for the diffusion of information, advertising, and marketing strategies in a social system extended on a surface.
Future research problems include  
 the competition of opinion
leaders in different states, the consideration of diverse interaction dynamics among agents, and the effects of opinion leaders in complex network topologies and in real social media.

\section*{Acknowledgments}

This work was supported by
Corporaci\'on Ecuatoriana para el Desarrollo de la
Investigaci\'on y Academia (CEDIA); through project
CEPRA-XIII-2019 ``Sistemas 
Complejos''.

\end{document}